\documentclass[preprint]{nrc1}

\usepackage{amsmath, amssymb, bm, cite, graphicx, graphics, color,mathrsfs}

\newcommand{\Lie}[0]{{\cal L}\, }

\newcommand{\tl}{\theta_{(\ell)}}
\newcommand{\tn}{\theta_{(n)}}

 \newcommand{\sgn}{\operatorname{sgn}}

\newcommand{\nn}{\nonumber}
\newcommand{\be}{\begin{equation}}
\newcommand{\ee}{\end{equation}}
\newcommand{\bea}{\begin{eqnarray}}
\newcommand{\eea}{\end{eqnarray}}

\newcommand{\tq}{\tilde{q}}
\newcommand{\hu}{\hat{u}}
\newcommand{\hn}{\hat{s}}

\newcommand{\cV}{\mathcal{V}}

\journal{ }

\begin{document}

\title{Two physical characteristics of numerical apparent horizons}
\author {Ivan Booth}
\address{Department of Mathematics and Statistics, Memorial
 University\\
St. John's, Newfoundland and Labrador, A1C 5S7, Canada \\
ibooth@math.mun.ca}
\shortauthor{I. Booth}

\maketitle

\begin{abstract}
This article translates some recent results on quasilocal horizons
into the language of $(3+1)$ general relativity so as to make them
more useful to numerical relativists. In particular quantities are
described which characterize how quickly an apparent 
horizon is evolving and how close it is to either equilibrium 
or extremality. 
\end{abstract}

\section{Introduction}

In $(3+1)$-dimensional numerical relativity an apparent horizon 
in a three-dimensional spacelike slice $\Sigma_t$ is the outermost
smooth two-surface whose whose outward null expansion vanishes
\cite{thomas,thornburg}. Then an apparent horizon world-tube 
is a three-surface $H$
foliated by two-surfaces $S_t \subset \Sigma_t$ with vanishing outward 
null expansion. 
Such surfaces have been an active topic of research in mathematical
relativity for the last 10-15 years (see \cite{AKRev,GourgRev,myRev} for 
reviews or the reference section of \cite{BigPaper} for more recent papers) 
and results from that work are now beginning 
to be applied to numerical relativity \cite{KrishNum1,KrishNum2}. 

This short paper continues this tradition by translating results from recent
papers into the language of $(3+1)$-dimensional relativity.  We 
first consider the recent work on slowly evolving horizons 
\cite{SEH,SEH2,BigPaper}. These papers found conditions under
which a horizon can be considered to be in a quasi-equilibrium
state. Intuitively, slowly evolving horizons
are ``almost" isolated \cite{iso} and so ``nearly" null.  Here we shall 
focus on an expansion parameter that arose from that
work. This parameter is necessarily small for slowly evolving horizons,
but even if it isn't small it can be used to invariantly characterize
the rate of expansion of horizons.  
It is to this end that we discuss it here, with a particular eye to 
tracking how quickly a black hole settles down to 
equilibrium after a merger or formation. 

Second we consider a parameter that characterizes how 
close an apparent horizon is to being extremal\cite{extremal}.
Extremality in this case is a generalization of the well-known bound
on the maximum angular momentum of a Kerr black hole relative to its
mass (or equivalently surface area). For apparent horizons this 
turns out to be  related to the expansion parameter and
places a bound on the maximum angular momentum relative to the
intrinsic geometry of the horizon. The immediate use
for this parameter would be to study the proximity to extremality of post-merger
black holes. 

\section{Horizons in (3+1)-dimensional gravity}

Let $(M,g_{ab}, \nabla_a)$ be a four-dimensional spacetime which 
is foliated by spacelike three-surfaces 
$\{ (\Sigma_t, h_{ij} , D_i) \,  ; t_1 \leq t \leq t_2\}$ where $h_{ij}$ is the induced metric and $D_i$ is the compatible covariant derivative. 
Note the indices; throughout this paper we will 
use early-alphabet latin indices for four-dimensional quantities 
and mid-alphabet latin indices for those defined in the three-slices. 
We use $e$ to denote the pull-back/push-forward operator
between the various spaces. 

Next let $T^a$ be a compatible time-evolution vector field
generating a flow which maps the $\Sigma_t$ into each other. That is 
$\Lie_T t = 1$. In the usual way this may be broken up as 
\bea
T^a = N \hat{u}^a + V^a \, , 
\eea
where $N$ is the lapse function, $V^a = e^a_i V^i$ 
is the spacelike shift vector field for some 
$V^i \in T \Sigma_t$, and $\hat{u}^a$ is the future-pointing 
timelike normal vector field to the $\Sigma_t$. 

Finally we choose our sign convention so that the extrinsic curvature $K_{ij}$ of the $\Sigma_t$ in $M$ is
\bea
K_{ij} = \frac{1}{N} ( \dot{h}_{ij} - \Lie_V h_{ij} ) 
= e_i^a e_j^b \nabla_a \hat{u}_b \, , 
\eea
where the dot indicates a Lie derivative with respect to $T^a$.

\begin{figure}[t]
\begin{center}
\scalebox{0.8}{\includegraphics{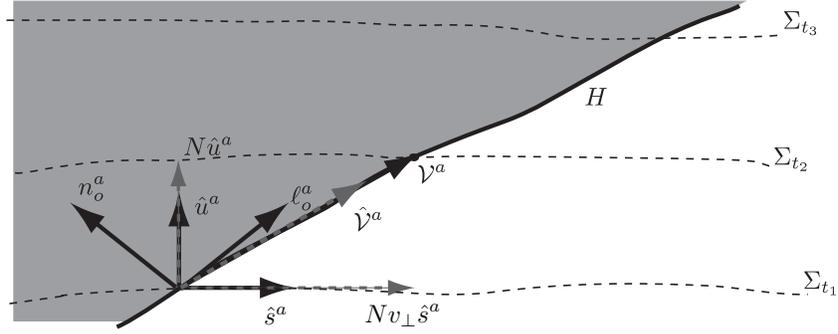}}
\caption{A two-dimensional cross-section of a spacetime $M$ foliated by spacelike three-surfaces $\Sigma_t$. $H$ is an apparent horizon world-tube which is foliated by the spacelike two-surfaces $S_t = H \cap \Sigma_t$. 
  }
\label{NormVect}
\end{center}
\end{figure}

Now, let $S_t$ be a closed two-surface in a slice $\Sigma_t$ and
write its outward-pointing unit normal as $\hat{s}^i$. Then 
future-outward and future-inward pointing null normals to 
$S_t$ in $M$ can be written as
\bea
\ell^a = f ( \hat{u}^a + \hat{s}^a )  \; \; \mbox{and} \; \; 
n^a = \frac{1}{2f } \left( \hat{u}^a - \hat{s}^a \right) \, , 
\label{scaling}
\eea
where $\hat{s}^a = e^a_i \hat{s}^i$ and $f$ is an arbitrary 
positive function: we have adapted the standard cross-normalization 
$\ell \cdot n = -1$ so there is the only free parameter in their definition. 
For $f=1$ we denote this pair of  null normals as 
$(\ell^a_o, n^a_o)$ as shown in Fig.~\ref{NormVect}. 

We write the induced two-metric on $S_t$ as $\tilde{q}_{AB}$ 
(using capital Latin indices for such two-dimensional quantities) 
and note that 
\bea
\tilde{q}^{ab} \equiv e^a_A e^b_B \tq^{AB}
= g^{ab} + \ell^a n^b + \ell^b n^a 
\; \; \mbox{and} \; \; 
\tilde{q}^{ij} \equiv e^i_A e^j_B \tq^{AB}
= h^{ij} - \hat{s}^i \hat{s}^j \, . 
\eea
Then the outward and inward null expansions are, respectively,
\bea
\tl = \tq^{ab} \nabla_a \ell_b = f (\tq^{ij} K_{ij} + \tq^{ij} D_i \hat{s}_j )
\; \; \mbox{and} \; \; 
\tn = \tq^{ab} \nabla_a n_b = \frac{1}{2f} (\tq^{ij} K_{ij} - \tq^{ij} D_i \hat{s}_j ) \, . 
\label{expansions}
\eea
As noted in the introduction, on a numerical apparent horizon 
$\tl = 0$ and in practical simulations where there are known to be
black holes in the initial data, this will usually be a sufficient 
condition to identify a black hole boundary. A key word in the 
previous sentence is ``usually" and we will return to the  
potential complications in section \ref{Sect_Extr}. 

For now, however, we will assume that an apparent horizon finding algorithm
\cite{thornburg} 
has been used to identify  $\tl = 0$ surfaces 
on each $\Sigma_t$ over some range of $t$ and 
further that over that range their union is a smooth three-surface 
$H$. Then we can always find a horizon evolution vector field
$\mathcal{V}^a$ that is normal to the $S_t$ and maps these surfaces
into each other. Since the $S_t \subset \Sigma_t$ we can write
\bea
\mathcal{V}^a = N (\hat{u}^a + v_\perp \hat{s}^a ) \, , 
\label{FoliationV}
\eea
where $N$ is again the lapse and $v_\perp$ is the the ``velocity"
of the horizon relative to the foliation. For a null horizon parallel to $\ell^a$, 
$v_\perp = 1$ while for a spacelike and expanding horizon $v_\perp>1$. 

\section{Expansion}
\label{Sect:Exp}

We now consider the first of the physical characteristics: the rate of
expansion of the horizon. Relative to the foliation, the rate 
of change of the area element $\sqrt{\tilde{q}}$ is 
\bea
\Lie_{\cV} \sqrt{\tq} = \sqrt{\tq} 
\left( \tilde{q}^{ab} \nabla_a \mathcal{V}_b \right) 
 =\sqrt{\tq}   \left( N (v_\perp - 1) \tilde{q}^{ij} D_i \hat{s}_j \right) \, , \nn
\eea
where we have made use of the fact that 
$\tl = 0 \Leftrightarrow  \tq^{ij} K_{ij} = - \tilde{q}^{ij} D_i \hat{s}_j$
(which could also be used to rewrite the last line in terms of
$\tq^{ij} K_{ij}$). 
This expansion is manifestly dependent on the lapse $N$ and so the
coordinate $t$ labelling of the hypersurfaces $\Sigma_t$. However, 
if $H$ is not null there is an obvious 
remedy to this problem : calculate the rate of expansion with respect to the unit 
$\hat{\cV}^a = \cV^a / || \cV ||$ instead of $\cV$ itself. Then the
expansion relative to $\hat{\cV}$ is
\bea
\theta_{(\hat{\cV})} = \alpha \sqrt{\left| \frac{v_\perp - 1}{v_\perp + 1} \right| } \tilde{q}^{ij} D_i \hat{s}_j \, ,
 \eea
where $\alpha = \sgn(v_\perp-1)$. 
Note that $\theta_{(\hat{\cV})}$ goes to zero as $v_\perp \rightarrow 1$ 
even though $\hat{\cV}^a$ itself diverges. 
%

To better understand $\theta_{(\hat{\cV})}$, we return to four-dimensions and 
the formalism developed for \emph{slowly evolving horizons} 
\cite{SEH,BigPaper}. 
There, $\cV^a$ was written in terms of the null normals and a 
parameter $C$:
\bea
\mathcal{V}^a = \ell^a - C n^a \, .
\label{HorizonV}
\eea
Then the foliation of the horizon fixes both a scaling of the null vectors
and $C$. Comparing with (\ref{FoliationV}) it is straightforward to see that in the 
$(3+1)$-formalism these take the form:
\bea
f = N (1 + v_\perp)/2 \; \; \mbox{and} \; \; 
C = {N^2}(v_\perp^2-1)/2 \, . 
\label{C}
\eea
The value of $C$ fixes the signature of the horizon: 
$C > 0 \Leftrightarrow |v_\perp| > 1 \Leftrightarrow$ $H$ is spacelike, 
$C = 0 \Leftrightarrow |v_\perp| = 1 \Leftrightarrow$ $H$ is null, and
$C < 0 \Leftrightarrow |v_\perp| < 1 \Leftrightarrow$ $H$ is timelike. 
We can rewrite
\bea
\theta_{(\hat{\cV})} = - \alpha \sqrt{|{C}/{2}|} \tn \, , 
\label{Cexp}
\eea
where $\alpha = \sgn(C)$. In this form
$\theta_{(\hat{\cV})}$ (or actually its square) is the most important of several
quantities used to decide whether or not a horizon is ``almost" isolated 
and so in quasi-equilibrium with its surroundings. Among other properties
such horizons will obey approximate zeroth law and first laws of black hole
mechanics with all fluxes being calculated relative to the $\ell^a$ normals (as if they 
where truly null). Some intuition for how dramatically a horizon can evolve and 
still be counted as in a quasi-equilibrium state can be gained 
from the examples of  \cite{SEH2}. 

To be mathematically certain about such a classification one needs to track
$\theta_{(\hat{\cV})}^2$ and the other quantities point-by-point on the horizons.
However to get a quick idea of the rate of expansion (and proximity to equilibrium)
it is convenient to have a single, dimensionless, number to calculate. Such a 
quantity can be obtained by integrating the square of $\theta_{(\hat{\cV})}$ 
over $S_t$:
\bea
\epsilon^2_{area} = \frac{1}{2} \int_{S_t}  \mspace{-10mu} d^2 x \sqrt{\tilde{q}} |C| \tn^2  
= \int_{S_t} \mspace{-10mu} d^2 x \sqrt{\tilde{q}}  
\left| \frac{v_\perp - 1}{v_\perp + 1} \right| (\tilde{q}^{ij} D_i \hat{s}_j)^2   \, .
\label{epsilon_area}
\eea
In spherical symmetry this is proportional to $(dR_H/ds)^2$, where
$R_H= \sqrt{a/4 \pi}$ is the areal radius of $S_t$ and $s$ is the arclength
``up" the horizon along a flow line of $\cV^a$. 

\section{Apparent horizons as black hole boundaries}
\label{Sect_Extr}

\begin{figure}[t]
\begin{center}
\scalebox{0.8}{\includegraphics{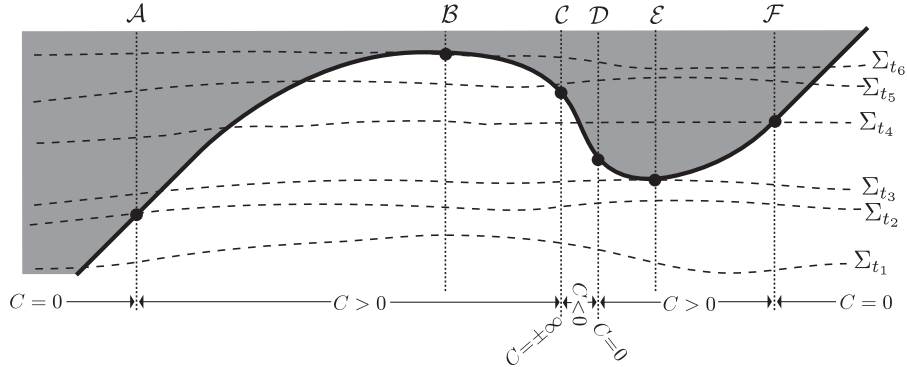}}
\caption{A horizon $H$ with $\tl = 0$ and $\tn < 0$
evolves between equilibrium states. Here, null horizons 
have slopes of $\pm 45^\circ$, spacelike
horizons have slopes from $-45^\circ$ to $45^\circ$, and timelike
membranes have the steeper slopes outside of that range. 
Thus before $\mathcal{A}$, $H$ is in equilibrium and null (parallel to $\ell^a$). 
From $\mathcal{A} \rightarrow \mathcal{C}$ it is expanding and spacelike. 
At $\mathcal{C}$ it is momentarily null (parallel to $n^a$) and $\epsilon_{area}$ 
diverges.  From $\mathcal{C} \rightarrow \mathcal{D}$, $H$ is a timelike membrane
before becoming null again at $\mathcal{D}$. 
Finally from $\mathcal{D} \rightarrow \mathcal{F}$ it is again a dynamical horizon
before becoming isolated and quiescent after $\mathcal{F}$. 
Between $\mathcal{C}$ and $\mathcal{D}$, $H$ is super-extremal
while at those points it is extremal. 
}
\label{HB}
\end{center}
\end{figure}

We now return to the ``usually" that followed
equation (\ref{expansions}) and consider when 
we can definitively associate a black hole with an apparent horizon. 
Following Penrose \cite{penrose} we take the existence of 
fully trapped surfaces ($\tl < 0$, $\tn < 0$) as being 
the key characteristic of black holes. Together with the null energy
condition (and some more technical assumptions) the existence
of trapped surfaces both implies a spacetime singularity somewhere
in their causal future and, if the spacetime is asymptotically flat, 
an enveloping event horizon  \cite{hawkellis}.  
Then, taking the set of all points that lie on trapped surfaces as
the interior of a black hole, one would intuitively expect the 
boundary of that region to be foliated by two-surfaces $S_t$
with $\tl = 0$ and $\tn < 0$, and further expect that slight inwards 
deformations of the $S_t$ should turn them into fully trapped 
surfaces. These ideas are the foundation for both the mathematical
relativity definition of an apparent horizon (the closure of the 
boundary of the set of all points lying on trapped surfaces in a given $\Sigma_t$) 
\cite{hawkellis}  
as well as Hayward's trapping horizons 
\cite{hayward}. 

Though in reality things turn out to be more complicated (see for example
the discussion of ``wild" trapped surfaces in \cite{eardley} or the counter
examples in \cite{bendov}) the motivation remains valuable and $\tl = 0$
``boundaries"  with trapped surfaces just inside are, at the very least, associated
with black holes and their boundaries. In preparation for discussing their
properties we quickly review deformations
(for more details see \cite{BigPaper}). A \emph{normal deformation}
 to a surface $S_t$
is characterized by a deforming vector field $X^a \in T_\perp S_t$ 
(the normal bundle to $S_t$). Then taking any smooth extension of $X^a$ into 
a neighbourhood of $S_t$ one can construct the corresponding flow and use
this to deform $S_t$; schematically $S_t \rightarrow S_t + \epsilon X$. The
deformation operator $\delta_X$ then calculates the rate of change of a quantity
during such an evolution. This operator is linear in the sense that
$\delta_{X+Y} = \delta_X + \delta_Y$ but in general $\delta_{fX} \neq f \delta_X$.

Thus, for an apparent horizon, the assumption that there be trapped surfaces 
``just inside" $S_t$ corresponds to there existing an inward pointing
vector field $X^a \in T{\Sigma_t}$ such that $\delta_X \tl < 0$. 
If such a two-surface exists on one slice $\Sigma_{t_o}$ then 
this two-dimensional apparent horizon can necessarily be extended into 
a three-dimensional, apparent horizon world-tube $H$ over some finite range
of $t$ \cite{ams}. If the null energy condition also holds, then  
$H$ is necessarily spacelike or null (that is $C \geq 0$) and non-decreasing
in area if $\tn < 0$  \cite{ams, hayward, ak} --- this last fact follows 
directly from 
$ \delta_{\cV} \sqrt{\tq} = - \sqrt{\tq} C \tn $. 
Null regions are \emph{isolated horizons} \cite{iso} and may be regarded as 
black hole equilibrium states while the spacelike and expanding regions are 
\emph{dynamical horizons} \cite{ak}. On any given 
$S_t$, $\cV^a$ will be either spacelike everywhere or null everywhere \cite{ams}. That
is, transitions between isolated and dynamical horizons must happen ``all at once":
no individual $S_t$ can be partly isolated and partly dynamical. 

It is possible for a horizon satisfying
$\tl = 0$, $\tn < 0$, and $\delta_n \tl < 0$ (note the use of the inwards null
direction for cases where there isn't a favoured inward spacelike direction)
to evolve into a 
\emph{timelike membrane} on which $\delta_n \tl > 0$ \cite{bendov2,mttpaper,thornburg}.
Such a situation is shown in Fig.~\ref{HB}. Physically, evolutions of this type 
correspond to situations where new horizons form outside of existing ones. 
The cited examples were for spherically symmetric spacetimes but it is widely 
speculated that something similar happens during at least some 
of the apparent horizon ``jumps" seen in numerical relativity.  For example,
if one was only tracking the outermost $\tl = 0$ surface  in Fig.~\ref{HB}, it would 
jump at $t = t_3$.

It is useful to compare $C$ and $v_\perp$ in this
figure. While the sign of $C$ faithfully tracks the signature of $H$, $v_\perp$
is coordinate dependent and diverges at $\mathcal{B}$ and $\mathcal{E}$
even though nothing physically significant happens there. 
For $v_\perp$, the physically
significant values are $\pm 1$ not zero or infinity. 

\section{Evolutions and extremality for dynamical horizons}

In the example depicted in Fig.~\ref{HB}, $C$ diverges and $H$ changes 
signature when $\delta_n \tl = 0$. In spherical symmetry $\delta_n \tl$ doesn't
depend on the scaling of the null vectors, but in general one must select a ``correct" 
scaling to observe this. For dynamical horizons this scaling is easy to find. 
Relative to the null vectors of section \ref{Sect:Exp},
rescale so that $\bar{\ell} = \ell/C$ and
$\bar{n} = C n$ or equivalently choose $\bar{f} = 1/(N(v_\perp - 1))$. Then
since $\tl = 0$ everywhere on $H$ we have \cite{BigPaper,extremal} 
\bea
 \delta_\cV \theta_{\bar{\ell}} = C \delta_{\bar{\ell}} \theta_{(\bar{\ell})} 
- \delta_{\bar{n}} \theta_{(\bar{\ell})} = 0  \; \; \mbox{where} \; \; 
\delta_{\bar{\ell}} \theta_{\bar{\ell}} = 
-\sigma_{(\bar{\ell})}^{AB}  \sigma_{(\bar{\ell})AB} -
 8 \pi G T_{ab} \bar{\ell}^a \bar{\ell}^b \, . 
\label{dvtl}
\eea
$\sigma_{(\bar{\ell}) AB} =  e_A^a e_B^b \nabla_a \bar{\ell}_b$ is the shear in the
$\bar{\ell}^a$ direction (note its simplified form thanks to $\theta_{(\bar{\ell})}= 0$). 
One of the reasons for the scaling choice is now clear:
for general deformations $\delta_{f X} \neq f \delta_X$ but
$\delta_\ell \tl$ is a special case where multiplicative factors can
be commuted through $\delta$. 
From this result it is easily seen that if the null energy condition holds and
$\delta_{\bar{\ell}} \theta_{(\bar{\ell})} \neq 0$ then 
$C > 0 \Leftrightarrow \delta_{\bar{n}} \theta_{(\bar{\ell})} < 0$ (and vice versa).
This is one proof of the already mentioned result that dynamical apparent horizons
with trapped surfaces just inside are generically spacelike. Further note that if 
$\delta_{\bar{n}} \theta_{(\bar{\ell})} \rightarrow 0$ then $C$ necessarily diverges. 

In \cite{extremal} it has been argued that the most natural way to 
define extremality for dynamical horizons is in terms of 
$\delta_{\bar{n}} \theta_{(\bar{\ell})}$\footnote{However in near-isolated cases where $C \approx 0$ and so the suggested rescaling $\bar{\ell} = \ell / C$ 
may be numerically inappropriate,
other methods of calculating this quantity may be more useful \cite{extremal}.}. A sub-extremal horizon is one 
for which $\delta_{\bar{n}} \theta_{\bar{\ell}} < 0$ 
(and so there are trapped surfaces just
inside) while for an extremal horizon $\delta_{\bar{n}} \theta_{\bar{\ell}} = 0$. 
This definition follows by extension from isolated horizons such as Kerr where it 
implies the usual restrictions on surface gravity (positive for sub-extremal,
vanishing for extremal) and, in cases where it is well-defined, angular
momentum. Note however that this characterization also applies
to highly  dynamical and distorted horizons where the angular momentum 
and surface gravity may not be so well-defined. 

For a dynamical numerical apparent horizon $\delta_{\bar{n}} \theta_{\bar{\ell}}$ can be calculated directly as it was in \cite{extremal} but in most cases it is probably 
easier to combine equations (\ref{C}) and (\ref{dvtl}) to obtain:
\bea
\delta_{\bar{n}} \theta_{(\bar{\ell})} 
= - \frac{1}{2} \left( \frac{v_\perp + 1}{v_\perp - 1} \right)
(  || \sigma_{(\ell_o)} ||^2 + 8 \pi G T_{ab} \ell_o^a \ell_o^b)  
\label{Extremal}
\eea
where $\ell_o^a = \hu^a + \hn^a$ and so 
$\sigma_{(\ell_o) AB} =  e_A^i e_B^j (K_{ij} + D_i \hn_j )$. 
Clearly
this expression goes to zero if $v_\perp \rightarrow -1$ ($H$ is null and 
parallel to $-n^a$) and becomes positive for timelike membranes.
%
As for the expansion we can define a dimensionless parameter that tracks how
close a horizon is to extremality. We call this the \emph{extremality parameter}:
\bea
e = 1 +\frac{1}{4 \pi} \int_{S_t} \mspace{-10mu} d^2 x 
\sqrt{\tilde{q}} \delta_{\bar{n}} \tl = 1 - \frac{1}{8 \pi} \int_{S_t}  \mspace{-10mu} d^2 x
 \sqrt{\tilde{q}} 
\left( \frac{v_\perp + 1}{v_\perp - 1} \right)
(  || \sigma_{(\ell_o)} ||^2 + 8 \pi G T_{ab} \ell_o^a \ell_o^b)  \, . 
\label{e}
\eea
If the dominant energy condition holds $e \geq 0$ (this is most easily seen
from the alternative method of calculating quantity described in \cite{extremal}). 
Thus, for sub-extremal holes $0 \leq e <1$ while for an extremal horizon
$e=1$. Super-extremal timelike membranes will have $e>1$. 
Note however that the intended use of this parameter is not to determine
the signature of the horizon but rather to quantify how close a horizon is to 
extremality and so a transition to a timelike membrane. If one is only interested
in signature, $v_\perp$ is sufficient. 
%


\section{Conclusions}
This short paper has defined two dimensionless quantities 
$\epsilon_{area}$ (Eq.~\ref{epsilon_area}) and $e$ (Eq.~\ref{e}) which can be used
to characterize dynamic black hole evolutions. $\epsilon_{area}$ invariantly defines the 
rate of expansion: it is very small for slowly evolving horizons and vanishes only for 
truly isolated horizons. As such it is well suited to tracking the approach of a dynamical 
horizon to equilibrium. By contrast $e$ is probably of greatest interest in highly 
dynamical situations. As noted $0 \leq e <1$ for regular sub-extremal holes, achieves
unity for extremal horizons and becomes greater than one for super-extremal timelike membranes. Thus it is well suited to tracking the approach to or aftermath of transitions 
between timelike membranes and dynamical horizons --- though it should be kept in mind
that $\epsilon_{area}$ also has something to say about such situations as it diverges
whenever $e = 1$.




\section*{Acknowledgements}
This work was supported by Natural Sciences and Engineering Research Council
of Canada.

\end{document}